\documentclass[prd,twocolumn,floats,floatfix,showpacs,nofootinbib]{revtex4}
\usepackage{graphicx}
\usepackage{dcolumn}
\usepackage{bm}
\usepackage{graphics}
\usepackage{slashed}
\usepackage{amssymb}
\usepackage{natbib}
\usepackage{amsmath}
\usepackage{url}
\usepackage{color}
\newcommand{\bea}{\begin{eqnarray}}
\newcommand{\ena}{\end{eqnarray}}
\newcommand{\bean}{\begin{eqnarray*}}
\newcommand{\enan}{\end{eqnarray*}}

\newcommand{\fracc}[2]{\frac{\textstyle{#1}}{\textstyle{#2}}}

\begin{document}

\title{Gaussian coordinate systems for the Kerr metric}

\author{M. Novello \footnote{M. Novello is Cesare Lattes ICRANet Professor}} \email{novello@cbpf.br}
\author{E. Bittencourt}\email{eduhsb@cbpf.br}

\affiliation{Instituto de Cosmologia Relatividade Astrofisica ICRA -
CBPF\\ Rua Dr. Xavier Sigaud, 150, CEP 22290-180, Rio de Janeiro,
Brazil}

\date{\today}

\begin{abstract}
We present the whole class of Gaussian coordinate systems for the
Kerr metric. This is achieved through the uses of the relationship
between Gaussian observers and the relativistic Hamilton-Jacobi
equation. We analyze the completeness of this coordinate system. In
the appendix we present the equivalent JEK formulation of General
Relativity -- the so-called quasi-Maxwellian equations -- which
acquires a simpler form in the Gaussian coordinate system. We show
how this set of equations can be used to obtain the internal metric
of the Schwazschild solution, as a simple example. We suggest that
this path can be followed to the search of the internal Kerr metric.
\end{abstract}

\maketitle

\section{Introduction}
The recognition that natural processes cannot be influenced by any
choice of a representation of the events occurring in spacetime
conducted the covariance principle to be assumed as one of the
fundamentals of modern physics. This idea was explicitly used to
build a theory of gravity by Einstein, General Relativity, which gave
a step forward by invoking that the MMG (\textit{ Manifold
Mapping Group}) should be taken as an invariance principle of
the theory. In practical uses however, one is obliged to select a
particular language by choosing a special coordinate system to
describe a given phenomenon. Among all the possible choices one can
make -- most of them dictated by symmetry of the given problem --
there is a very special one that can bring us more physical
insight about the problem to be treated, that is the
\textit{Gaussian coordinate system} or \textit{Synchronic coordinate
system}. In fact, this coordinate system was suggested by C. Gauss in his works about
curves and surfaces. In Gaussian coordinate system (GCS), a foliation of
spacetime is made in such a way that one separates space and
time, as in pre-relativistic theories. The time-like world-line of an observer
that is orthogonal to the 3-d space (which is identified to the co-moving system)
is such that the proper time of such observer coincides with the coordinate time.

In the standard procedure made by the father founders of
relativistic cosmology, the Gaussian coordinate system appears
closely related to the Cosmological Principle: the large scale structure of the Universe
behaves as being homogeneous and isotropic. The solutions of Einstein equations which satisfies
this postulate possess a complete Gaussian coordinate system, in addition the Gaussian surface is a
Cauchy surface.

On the other hand, there are cosmological solutions which do not
satisfy this postulate. However these solutions have undesirable
properties, as for instance, closed timelike curves.

A similar approach was done for G\"odel's metric by one of us
\cite{guimaraes}. It was shown that the Gaussian coordinate system
is limited and cannot be extended beyond a certain region, the
domain of which depends only on the vorticity present in this
geometry. Such inaccessible region defines a frontier for time-like
geodesics, which prohibits the extension of the GCS into the whole
manifold.

Mathematically, a Gaussian coordinate system is constructed by the definition of a hypersurface $S=S(x^{\mu})$,
which satisfies

\begin{equation}
\label{eq_gauss}
\begin{array}{l}
g^{\mu\nu}\fracc{\partial S}{\partial x^{\mu}}\fracc{\partial S}{\partial x^{\nu}}=1,\\[2ex]
g^{\mu\nu}\fracc{\partial S}{\partial x^{\mu}}\fracc{\partial\bar
x^i}{\partial x^{\nu}}=0,
\end{array}
\end{equation}
where $\bar x^i$ are coordinates lying on $S$. The Gaussian
coordinates are given by $\bar x^{\mu}=(S,\bar x^i)$. Eq.\
(\ref{eq_gauss}) imposes $\bar g^{00}=1$ and $\bar g^{0i}=0$.

The main purpose of this work is to exhibit a Gaussian coordinate
system for the Kerr metric. The method we use is provided by the
relativistic Hamilton-Jacobi formalism of canonical transformations.
The key idea is to identify the principal Hamilton function with the
proper time of a test particle in this geometry. A immediate
application of this method can be done for other solutions of
Einstein equation as Schwarzschild and Kerr-Newman. Finally, in
appendix \ref{appendixA} we list the coordinate systems encountered
in literature for Kerr metric including our Gaussian coordinate
system and in appendix \ref{appendixB} we regain from the
\textit{Quasi-Maxwellian formalism} (JEK equations together with the
evolution equations of the kinematical quantities) the Schwarzschild
solution in a Gaussian coordinate system. This formalism is an
 equivalent way to obtain the results of General Relativity.


\section{Canonical Transformations Formalism}
The action for a free test particle with mass $m$ in the presence of a gravitational field can be written as

\begin{equation}
\label{action_free_part}
S=-m\int ds.
\end{equation}
Assuming that only one extreme of the path is fixed and the particle is moving on a geodesic line,
we make the variational principle and discover how $S$ depends on the coordinates $x^{\mu}$.
Such functional form is

\begin{equation}
\label{func_var_acao_part_livre}
\delta S=-mu_{\alpha}\delta x^{\alpha},
\end{equation}
where $u_{\alpha}$ is the 4-velocity. Defining the 4-momentum

\begin{equation}
\label{quadri_momentum} p_{\alpha}\doteq-\fracc{\partial S}{\partial
x^{\alpha}},
\end{equation}
it satisfies

\begin{equation}
\label{quadri_momentum_square}
p_{\alpha}p^{\alpha}=m^2.
\end{equation}

Replacing Eq.\ (\ref{quadri_momentum}) into\ (\ref{quadri_momentum_square}), we find the relativistic
Hamilton-Jacobi equation for a test particle in a gravitational field given by

\begin{equation}
\label{eq_ham_jac} g^{\mu\nu}\fracc{\partial S}{\partial
x^{\mu}}\fracc{\partial S}{\partial x^{\nu}}-m^2=0.
\end{equation}


\section{Equivalence between a Gaussian system and the Hamilton-Jacobi equation}

By a canonical transformation of coordinates of a physical system, so that $(q,p)$ goes to $(Q,P)$ via a
generating function $F_1(q,Q,t)$, we have the following constraint

\begin{equation}
\label{trans_can}
dF_1=\sum_i(p_idq_i-P_idQ_i)+(K-H)dt.
\end{equation}
In the covariant form, we write it as

\begin{equation}
\label{trans_can_cov}
dF_1(q^{\mu},Q^{\mu})=-p_{\alpha}dq^{\alpha}+P_{\mu}dQ^{\mu},
\end{equation}
where $p_{\alpha}=(H,-\vec p)$ and $P_{\mu}=(K\partial t/\partial Q^0,-\vec P)$. Using another generating function $F_2(q^{\mu},P_{\nu})$, such that

\begin{equation}
\label{func_gerat}
F_2=F_1-P_{\mu}Q^{\mu},
\end{equation}
we have

\begin{equation}
dF_2=-p_{\alpha}dq^{\alpha}-Q^{\mu}dP_{\mu}.
\end{equation}
Then the function $F_2$ must satisfy

\begin{equation}
\label{eq_f2}
\begin{array}{lcl}
p_{\alpha}&=&-\fracc{\partial F_2}{\partial q^{\alpha}},\\[2ex]
Q^{\mu}&=&-\fracc{\partial F_2}{\partial P_{\mu}}.
\end{array}
\end{equation}
Comparing with Eq.\ (\ref{quadri_momentum}), we identify $F_2(q,P)$ as the action $S(q,P)=S(q,p(q,P))$ of a test particle in a gravitational field. Therefore,
\begin{subequations}
\label{constr_s}
\begin{eqnarray}
p_{\alpha}&=&-\fracc{\partial S}{\partial q^{\alpha}},\label{constr_s1}\\[2ex]
Q^{\mu}&=&-\fracc{\partial S}{\partial P_{\mu}}.\label{constr_s2}
\end{eqnarray}
\end{subequations}
Assuming $Q^0=-S/m$, from the $0$-component of Eq.\ (\ref{constr_s2}), we get

\begin{equation}
\label{eq_q_0} \fracc{S}{m}=\fracc{\partial S}{\partial P_0},
\end{equation}
and then,

\begin{equation}
\label{sol_eq_q_0}
S=e^{P_0/m}f(q^{\mu},P_i).
\end{equation}
The Hamilton-Jacobi equation can be obtained making the new Hamiltonian $K$ constant ($H\rightarrow K\equiv const$) and,
as we know that $P_0\propto K$, it is possible to incorporate this constant into $S$ and it follows that $S=S(q^{\mu},P_i)$.
Hereupon, rewriting Eq.\ (\ref{eq_ham_jac}) we obtain

\begin{equation}
\label{eq_sist_gauss_1} g^{\mu\nu}\fracc{\partial Q^0}{\partial
x^{\mu}}\fracc{\partial Q^0}{\partial x^{\nu}}=1.
\end{equation}

We derive partially Eq.\ (\ref{eq_sist_gauss_1}) with respect to $P_i$ and use Eq.\ (\ref{constr_s2}) to get

\begin{equation}
\label{eq_sist_gauss_2} g^{\mu\nu}\fracc{\partial Q^0}{\partial
x^{\mu}}\fracc{\partial Q^i}{\partial x^{\nu}}=0.
\end{equation}
Note that the coordinate system $Q^{\mu}$ together with Eqs.\ (\ref{eq_sist_gauss_1}) and (\ref{eq_sist_gauss_2})
define a Gaussian coordinate system.


\section{Applications}

\subsection{Gaussian system for the Schwarzschild metric}

As a simple exercise, we exhibit a Gaussian coordinate system found
 from the definition\ (\ref{eq_gauss}) for internal and external Schwarzschild solution.
 This case is particularly simple due
to the symmetries of such metric.

\subsubsection{The external case}

Considering Schwarzschild external geometry described in the usual
coordinate system, for radial observers with non null velocity at
the infinity \cite{mario}, the metric which is originally given by

\begin{equation}
\label{schwarz}
ds^2=\left(1-\fracc{r_H}{r}\right)dt^2-\left(1-\fracc{r_H}{r}\right)^{-1}dr^2-r^2d\Omega^2,
\end{equation}
where $r_H=2M$ and $M$ is the geometrical mass of the gravitational source, becomes

\begin{equation}
\label{ds2_schwarz_mario_gauss}
ds^2=d\tau^2-\left(\alpha^2-1+\fracc{r_H}{r}\right)dR^2-r(\tau,R)^2d\Omega^2,
\end{equation}
according to the following coordinate transformation

\begin{equation}
\label{transf_schwarz_mario_gauss}
\left\{\begin{array}{lcl}
\tau=\alpha t+F_{e}(r,\alpha),\\
R\doteq\fracc{\partial\tau}{\partial\alpha},
\end{array}\right.
\end{equation}
where $\alpha\in\Re$ is a external parameter and $\tau$ is
interpreted as the proper time. Substituting this proposal in Eq.\
(\ref{eq_gauss}) one can see that $F'_{e}(r,\alpha)\equiv dF_{e}/dr$
must satisfy

\begin{equation}
\label{eq_f_r}
F'_{e}(r,\alpha)=\sqrt{\fracc{\alpha^2-(1-\fracc{r_H}{r})}{(1-\fracc{r_H}{r})^2}},
\end{equation}
for the new coordinate system to be admissible.

Analyzing the particular case in which the velocity is zero\
\cite{chineses}, we obtain the coordinate transformation integrating
the following geodesic equations parameterized by the proper time
$\tau$

\begin{equation}
\label{schwar_geo_rad}
\left\{\begin{array}{l}
\fracc{d^2t}{d\tau^2}+\fracc{A'}{A}\fracc{dt}{d\tau}\fracc{dr}{d\tau}=0,\\[2ex]
\fracc{d^2r}{d\tau^2}+\fracc{1}{2}A'A\left(\fracc{dt}{d\tau}\right)^2-\fracc{1}{2}\fracc{A'}{A}\left(\fracc{dr}{d\tau}\right)^2=0,\\[2ex]
\fracc{d\theta}{d\tau}=0,\\[2ex]
\fracc{d\phi}{d\tau}=0,
\end{array}\right.
\end{equation}
where $A=(1-r_H/r)$ and we obtain

\begin{equation}
\left\{\label{transf_schwarz_chin_gauss}
\begin{array}{lcl}
t&=&\tau+r_H\left[\ln\left(\fracc{\sqrt{r/r_H}+1}{|\sqrt{r/r_H}-1|}\right)-2\sqrt{\fracc{r}{r_H}}\right],\\[2ex]
r&=&\left[-\fracc{3}{2}\sqrt{r_H}(\tau+R)\right]^{2/3}.
\end{array}\right.
\end{equation}
The new metric is obtained just choosing $\alpha^2=1$ in Eq.\
(\ref{ds2_schwarz_mario_gauss}). It means that we are describing
particles whose mechanical energy is equal to the rest energy. From
this we conclude that $\alpha^2\geq1$. Note that this coordinate
transformation does not converge to the inverse coordinate
transformation of\ (\ref{transf_schwarz_mario_gauss}) if we choose
$\alpha^2\rightarrow1$ before the calculations, because we lose the
degree of freedom necessary to build the other coordinates. In both
cases we observe that the horizon ``disappears'', i.e., this
coordinate system does not have any problem for $r=r_H$ as in
Eddington-Finkelstein or Kruskal-Szekeres coordinates system.
However, in these cases, the natural "observers" are null-type. On
the other hand, in the Gaussian systems\
(\ref{transf_schwarz_mario_gauss}) and\
(\ref{transf_schwarz_chin_gauss}) the true natural observers
$V^{\mu}=\delta^{\mu}_0$ are time-like in every point, making
possible the description of events of the spacetime from geodetic
massive particles at rest.

\subsubsection{The internal case}\label{schwar_int_case}

To construct a Gaussian system for the interior solution, we can
consider a spherical shell filled with a perfect fluid of energy
density $\rho\equiv const.$ and pressure $p=0$ comoving to
$u_{\mu}=(e^{\nu/2},0,0,0).$ In this case the line element is given
by

\begin{equation}
\label{ds2_schwarz_int}
ds^2=e^{\nu(r)}dt^2-\left(1-\fracc{r_H}{r}\right)^{-1}dr^2-r^2d\Omega^2,
\end{equation}
where

\begin{equation}
e^{\nu(r)}=\left(\fracc{3}{2}\sqrt{1-\fracc{r_0^2}{r_c^2}}-\fracc{1}{2}\sqrt{1-\fracc{r^2}{r_c^2}}\right)^2,
\end{equation}
and $r_0$ is the radius of the ``star" and $r_c=3/\rho$. We also
assume that $r_0<r_c$ to avoid singulatities at the coordinate
system. Similar to the coordinate transformation\
(\ref{transf_schwarz_mario_gauss}), we have

\begin{equation}
\label{transf_schwarz_int_gauss}
\left\{\begin{array}{lcl}
\tau=\alpha t+F_{i}(r,\alpha),\\
R=\fracc{\partial\tau}{\partial\alpha},
\end{array}\right.
\end{equation}
where $F'_{i}(r,\alpha)\equiv dF_{i}/dr$ must satisfy

\begin{equation}
\label{eq_f_i_r}
F'_{i}(r,\alpha)=\sqrt{(\alpha^2e^{-\nu}-1)\left(1-\fracc{r_H}{r}\right)}.
\end{equation}

Therefore, the new line element for Schwarzschild interior solution, written in Gaussian coordinates, is

\begin{equation}
\label{ds2_schwarz_int_gauss}
ds^2=d\tau^2-(\alpha^2-e^{\nu})dR^2-r(\tau,R)^2d\Omega^2.
\end{equation}

Of course, at $r=r_0$ Eqs.\ (\ref{schwarz}) and\
(\ref{ds2_schwarz_int}) must be the same. Then we obtain that
$r_H=r_0^3/r_c^2$. The necessary continuity condition of the metric
on the star shell, in Schwarzschild coordinates, is naturally
carried to the Gaussian coordinates.


\subsection{Kerr Metric}\label{kerr_gauss}
In 1963 Kerr found an exact solution \cite{kerr} of Einstein
equations, which describes the external space-time generated by a
``source" having geometrical mass $M$ and angular momentum $a$ per
unit of geometrical mass. In 1968, Carter developed a method to
obtain all geodesics for this metric using the relativistic
Hamilton-Jacobi equation \cite{carter}. In this paper, he considers
 the principal Hamilton function $S$ as a function of the proper time
$\tau$ of a test particle. Here we identify both.

In the appendix we exhibit (see \cite{visser}), some known
coordinate systems of the Kerr metric and their main
characteristics. We include in this list the Gaussian system
presented here.

The line element for Kerr solution in Boyer-Lindquist coordinates is given by

\begin{equation}
\begin{array}{l}
\label{ds2kerr}
ds^2=\left(1-\fracc{2Mr}{\rho^2}\right)dt^2-\fracc{\rho^2}{\Delta}dr^2-\rho^2d\theta^2+\\[2ex]
+\fracc{4Mra\sin^2\theta}{\rho^2}dtd\phi+\\[2ex]
-\left[(r^2+a^2)\sin^2\theta+\fracc{2Mra^2\sin^4\theta}{\rho^2}\right]d\phi^2,
\end{array}
\end{equation}
where $\rho^2=r^2+a^2\cos^2\theta$ and $\Delta=r^2+a^2-2Mr$.

A simple way to calculate the geodesic equations parameterized by
the proper time for Kerr solution is to use the Euler-Langrange
equations found in \cite{chandrasekhar}. From first integrals of
these equations we construct the tangent vector field

\begin{equation}
\label{tan_vect_k}
\begin{array}{l}
V^{0}=\fracc{1}{\rho^2\Delta}[\Sigma^2E-2MraL],\\[2ex]
V^{1}=\fracc{F_n(r)}{\rho^2}\Delta,\\[2ex]
V^{2}=\fracc{G_n(\theta)}{\rho^2},\\[2ex]
V^{3}=\fracc{1}{\rho^2\Delta}[2Mr(aE-L\csc^2\theta)+\rho^2L\csc^2\theta],
\end{array}
\end{equation}
where $\Sigma^2=(r^2+a^2)^2-a^2\Delta\sin^2\theta$. The functions $F_n(r)$ and $G_n(\theta)$ are given by

\begin{equation}
\label{eq_f_g_kn}
\begin{array}{l}
F_n(r)=\pm\fracc{\sqrt{(E\tilde r^2-aL)^2-\Delta(r^2+\gamma)}}{\Delta},\\[2ex]
G_n(\theta)=\pm\sqrt{\gamma-a^2\cos^2\theta-\left(Ea\sin\theta-\fracc{L}{\sin\theta}\right)^2},
\end{array}
\end{equation}
where $\tilde r^2=r^2+a^2$ and $E$, $L$ and $\gamma$ are constants of integration.

Let us suppose that the new time coordinate is represented by $S$, a principal Hamilton function and written as follows

\begin{equation}
\label{ansatz_s_k}
S=-Et+L\phi+W_1(r)+W_2(\theta)\equiv-mQ^0.
\end{equation}
Substituting Eq.\ (\ref{ansatz_s_k}) in Eq.\ (\ref{eq_sist_gauss_1}), for $m=1$, we obtain

\begin{equation}
\label{eq_ham_jac_sub_1_k}
\begin{array}{l}
E^2\fracc{\Sigma^2}{\rho^2\Delta}-\fracc{\Delta}{\rho^2}\left(\fracc{d W_1}{dr}\right)^2-\fracc{1}{\rho^2}\left(\fracc{dW_2}{d\theta}\right)^2+\\[2ex]
-2\fracc{(2Mr-q^2)}{\rho^2\Delta}EL-\fracc{\Delta-a^2\sin^2\theta}{\rho^2\Delta\sin^2\theta}L^2=1,
\end{array}
\end{equation}
where $E$ is interpreted as being the energy and $L$ as being the angular momentum for a test particle when we consider the asymptotically flat regime. We can rewrite this expressionin a more convenient form like that

\begin{equation}
\label{eq_ham_jac_sub_2_k}
\begin{array}{l}
-\Delta\left(\fracc{dW_1}{dr}\right)^2+\fracc{(E\tilde r^2-aL)^2}{\Delta}-r^2=\\[2ex]
=\left(\fracc{dW_2}{d\theta}\right)^2+\left(Ea\sin\theta-\fracc{L}{\sin\theta}\right)^2+a^2\cos^2\theta\equiv\gamma,
\end{array}
\end{equation}
where $\gamma$ is a constant (of separability). So, we have two equations, one for $W^{'}_1$ and other for $W^{'}_2$ as

\begin{equation}
\label{eq_w_1_w_2_k}
\begin{array}{l}
\fracc{W_1}{dr}=\pm\fracc{\sqrt{(E\tilde r^2-aL)^2-\Delta(r^2+\gamma)}}{\Delta},\\[2ex]
\fracc{W_2}{d\theta}=\pm\sqrt{\gamma-a^2\cos^2\theta-\left(Ea\sin\theta-\fracc{L}{\sin\theta}\right)^2}.
\end{array}
\end{equation}
Note that $dW_1/dr=F_n(r)$ and $dW_2/d\theta=G_n(\theta)$.

The other coordinates, which we call here $Q^1\doteq R$, $Q^2\doteq\Theta$ and $Q^3\doteq\Phi$, are calculated from the spatial components of the second equation of\ (\ref{constr_s}) by

\begin{equation}
\label{other_coord_k}
\left\{\begin{array}{lcl}
R&\doteq&-\fracc{\partial S}{\partial E}=t-\fracc{\partial W_1}{\partial E}-\fracc{\partial W_2}{\partial E},\\[2ex]
\Theta&\doteq&-\fracc{\partial S}{\partial L}=-\phi-\fracc{\partial W_1}{\partial L}-\fracc{\partial W_2}{\partial L},\\[2ex]
\Phi&\doteq&-\fracc{\partial S}{\partial\gamma}=-\fracc{\partial
W_1}{\partial\gamma}-\fracc{\partial W_2}{\partial\gamma}.
\end{array}\right.
\end{equation}

The remaining metric components can be given in the form

\begin{equation}
\label{comp_metr_kerr_cov}
\begin{array}{lcl}
\bar g_{11}&=&1-\fracc{2Mr}{\rho^2}-E^2,\\[2ex]
\bar g_{12}&=&-EL-\fracc{2Mar\sin^2\theta}{\rho^2},\\[2ex]
\bar g_{13}&=&-2[E(\gamma-a^2)+aL],\\[2ex]
\bar g_{22}&=&-(L^2+\tilde r^2\sin^2\theta)-\fracc{2Mra^2\sin^4\theta}{\rho^2},\\[2ex]
\bar g_{23}&=&-2\left[L(r^2+\gamma)-(E\tilde r^2-aL)a\sin^2\theta\right],\\[2ex]
\bar g_{33}&=&-4(r^2+\gamma)(\gamma-a^2\cos^2\theta).
\end{array}
\end{equation}

The determinant of this new metric $\bar g_{\mu\nu}$ is

\begin{equation}
\label{det_bar_g}
\bar g\doteq\det\bar g_{\mu\nu}=-4f(r)h(\theta),
\end{equation}
where $f(r)=(E\tilde r^2-aL)^2-\Delta(r^2+\gamma)$ and
$h(\theta)=(\gamma-a^2\cos^2\theta)\sin^2\theta-(Ea\sin^2\theta-L)^2$.
If we guarantee that $f,h>0$ in some region of Kerr spacetime, then
the metric $\bar g_{\mu\nu}$ will be well-defined for these events.

The usual coordinate systems for Kerr metric present problems in
some regions, which are called \textit{horizons}. In Boyer-Lindquist
coordinates, for example, the mathematical expression for the Kerr
horizons are

\begin{equation}
\label{horiz_expr_kerr}
r_{\pm}=M\pm\sqrt{M^2-a^2}
\end{equation}
where $r_{+(-)}$ is called outer (inner) horizon. Differently, our
Gaussian system presents a complete regularity at the metric
components in the horizons. Note that there exists a divergence at
the real singularity $r=0$ and $\theta=\pi/2$.

We choose a $\bar V^{\mu}\doteq\delta^{\mu}_0$ in this coordinate
system, which is obviously geodetic. Taking the inverse coordinate
transformation, we write the observers field in Kerr coordinates in
such way that

\begin{equation}
\label{V_mu_k_coord}
\begin{array}{l}
V^{0}=\fracc{1}{\rho^2\Delta}[\Sigma^2E-2MraL],\\[2ex]
V^{1}=\fracc{W'_1}{\rho^2}\Delta,\\[2ex]
V^{2}=\fracc{W'_2}{\rho^2},\\[2ex]
V^{3}=\fracc{1}{\rho^2\Delta}[2Mr(aE-L\csc^2\theta)+\rho^2L\csc^2\theta].
\end{array}
\end{equation}

If we compare Eq.\ (\ref{V_mu_k_coord}) with Eq.\ (\ref{tan_vect_k}) obtained from
the Euler-Langrange equations we conclude that they are the same. Therefore, $\delta^{\mu}_0$ corresponds to
all tangent vectors of timelike geodesics for Kerr solution.\\

\section{Completeness of the Gaussian coordinate system}
At present there is not a theorem that specifies the necessary and
sufficient conditions of existence of a complete Gaussian coordinate
system for an arbitrary metric or whether the hyper-surface defined
by it is a Cauchy surface \cite{hawking}.

In our case, it is obvious the impossibility to identify $S$ with a
Cauchy surface due to the presence of closed time-like curves. What
about the completeness of such Gaussian system? To answer this
question one can analyze eventual divergences in the expansion
factor  $\vartheta.$ This will allow us to recognize the regions
where a given congruence of observers remains well-defined or not. A
given congruence of observers is nothing but a given choice of the
parameters $E$, $L$ and $\gamma.$  The explicit expression of the
expansion is

\begin{equation}
\label{expan_k_g}
\begin{array}{l}
\vartheta=\fracc{\cos\theta}{\rho^2\sqrt{h}}\big[\gamma-a^2\big(\cos^2\theta+(2E^2-1)\sin^2\theta\big)+2EaL\big]+\\[2ex]
-\fracc{1}{\rho^2\sqrt{f}}\big[2Er(E\tilde r^2-aL)-r\Delta-(r^2+\gamma)(r-M)\big],
\end{array}
\end{equation}
where $f$ and $h$ were defined at Sec.\ [\ref{kerr_gauss}]. So at the singularity $r=0$ and $\theta=\pi/2$ we obtain

\begin{equation}
\label{expan_sing_k_g}
\vartheta_s=\lim_{\theta\rightarrow\pi/2}\fracc{[{\cal Q}-a^2(E^2-1)+L^2]}{a^2\cos\theta\sqrt{{\cal Q}}}-\lim_{r\rightarrow0}\fracc{\gamma M}{ar^2\sqrt{-{\cal Q}}},
\end{equation}
where ${\cal Q}\doteq\gamma-(Ea-L)^2$. We conclude that the divergence $\vartheta\rightarrow -\infty$ at the singularity is guaranteed just for the congruence which have ${\cal Q}=0$. All the other congruences cannot reach the singularity.

If we pick out a given black hole with $M=2$ and $a^2=1$ (fixed that $M^2>a^2$, these numbers are completely arbitrary and they do not interfere at the results), we can make a more detailed analysis controlling the boundaries of the functions $f(r)$ and $h(\theta)$. Substituting $x=\sin^2\theta$ in $h(\theta)$, we get

\begin{equation}
\label{h_kerr}
h(x)=(1-E^2)x^2+(\gamma-1+2EL)x-L^2.
\end{equation}

By the roots of this polynomial function, we can encounter intervals
of the parameters range such that we can cover all values of
$\theta$. Thus, we can choose $L=0$ and $E^2>1$, and consequently we
obtain $E^2\leq\gamma$. By the other hand the $f(r)$ function
becomes

\begin{equation}
\label{f_kerr}
f(r)=(E^2-1)r^4+4r^3+(2E^2-1-\gamma)r^2+4\gamma r+(E^2-\gamma).
\end{equation}
From the zero-order term in $r$, we see that $E^2\geq\gamma$ is a
necessary condition for $f(r)$ to be greater than $0$ for all values
of $r$. Therefore, we conclude that is impossible to cover all
manifold events with only one congruence of observers. The case
${\cal Q}=0$ provides the only congruence which can cover all values
of $\theta$ and all positive values of $r$.


\section{Generalization (Kerr-Newman Metric)}
In 1965 E. Newman et al. \cite{newman} found a generalization of
Kerr solution, which describes a black hole with geometrical mass
$M$, angular momentum $a$ per unit of geometrical mass and charge
$q$.

The line element for Kerr-Newman solution is

\begin{equation}
\begin{array}{l}
\label{ds2kerr_newman}
ds^2=\left(1-\fracc{(2Mr-q^2)}{\rho^2}\right)dt^2-\fracc{\rho^2}{\Delta}dr^2-\rho^2d\theta^2+\\[2ex]
+\fracc{2(2Mr-q^2)a\sin^2\theta}{\rho^2}dtd\phi+\\[2ex]
-\left[(r^2+a^2)\sin^2\theta+\fracc{(2Mr-q^2)a^2\sin^4\theta}{\rho^2}\right]d\phi^2,
\end{array}
\end{equation}
where $\rho^2=r^2+a^2\cos^2\theta$ and $\Delta=r^2+a^2+q^2-2Mr$.

If we make all steps used in the previous section, at the end we find the covariant metric components for Kerr-Newman solution described by a Gaussian coordinate system, which can be given as follows

\begin{equation}
\label{comp_metr_kn_cov}
\begin{array}{lcl}
\bar g_{11}&=&1-\fracc{2Mr-q^2}{\rho^2}-E^2,\\[2ex]
\bar g_{12}&=&-EL-\fracc{(2Mr-q^2)a\sin^2\theta}{\rho^2},\\[2ex]
\bar g_{13}&=&-2[E(\gamma-a^2)+aL],\\[2ex]
\bar g_{22}&=&-(L^2+\tilde r^2\sin^2\theta)-\fracc{(2Mr-q^2)a^2\sin^4\theta}{\rho^2},\\[2ex]
\bar g_{23}&=&-2\left[L(r^2+\gamma)-(E\tilde r^2-aL)a\sin^2\theta\right],\\[2ex]
\bar g_{33}&=&-4(r^2+\gamma)(\gamma-a^2\cos^2\theta).
\end{array}
\end{equation}

Now $\bar V^{\mu}\doteq\delta^{\mu}_0$ corresponds to all tangent vectors of timelike geodesics for Kerr-Newman solution.\\

If we take appropriate limits we get Gaussian coordinate systems for other metrics: in the case $a=0$ and $q\neq0$ we obtain a Gaussian system for Reissner-Nordström solution, if $a\neq0$ and $q=0$ we construct the previous Gaussian system for the Kerr metric, and finally, if $a=q=0$ we obtain the Gaussian system for Schwarzschild case.


\section{Conclusions}
Following a systematic algorithm, we have built a set of Gaussian
coordinate systems for the Kerr metric and generalizations. We
analyzed the completeness of the Gaussian system which is
intrinsically related to the conserved quantities associated to a
test observer immersed in this geometry. Another important feature
is the fact that in this coordinate system the K-metric is not
static, i.e., a Gaussian observer measuring the geometric properties
in its neighborhoods, concludes that such metric is not static
according to its own proper time. In the appendix we show how it is
possible to use such GCS in order to undertake the search of the
corresponding internal solution. This work is under analysis.

\subsection{Acknowledgments}
E.B. would like to thank CNPq for the financial support. M.N thanks CNPq and FAPERJ.


\appendix
\section{Some Special coordinate systems for the Kerr metric}\label{appendixA}
As we said, in \cite{visser} we see many coordinate systems for Kerr metric encountered in literature and the coordinate transformations between them. Here there is a list of them, with some properties, and the inclusion of our Gaussian system.

\begin{itemize}

\item{Kerr's original coordinates:}

In this coordinate system $(u,r,\theta,\phi)$, the line element is

\begin{equation}
\begin{array}{l}
\label{ds2kerr_kr}
ds^2=\left(1-\fracc{2Mr}{\rho^2}\right)du^2-\rho^2d\theta^2-2a\sin^2\theta dr d\phi\\[1ex]
-2dudr-\fracc{4Mra\sin^2\theta}{\rho^2}dud\phi+\\[1ex]
-\left[(r^2+a^2)\sin^2\theta
+\fracc{2Mra^2\sin^4\theta}{\rho^2}\right]d\phi^2,
\end{array}
\end{equation}
where $\rho^2\equiv r^2+a^2\cos^2\theta$.

The most important features in this coordinate system are:

\begin{itemize}

\item{The appearance of a actual singularity in $g_{uu}$, for $r=0$ and $\theta=\pi/2$;}

\item{Setting $a\rightarrow0$, the line element reduces to Schwarzschild geometry in ``advanced Eddington-Finkelstein coordinates";}

\item{In terms of $M$, the line element can be put into Kerr-Schild form by $g^{\mu\nu}=g_0^{\mu\nu}+f(M,a,x^{\alpha})l^{\mu}l^{\nu}$, where $g_0^{\mu\nu}$ is Minkowski spacetime and $l^{\mu}$ é a geodetic null vector.}

\end{itemize}

\item{Kerr-Schild ``Cartesian" coordinates:}

By use of $(x^0,x,y,z)$ coordinates, the $ds^2$ for Kerr metric becomes

\begin{equation}
\label{ds2kerr_mink}
\begin{array}{l}
ds^2=(dx^0)^2-dx^2-dy^2-dz^2+\\[2ex]
-\fracc{2Mr^3}{r^4+a^2z^2}\left[dx^0+\fracc{r}{a^2+r^2}(x\,dx+y\,dy)+\right.\\[2ex]
\left.+\fracc{a}{a^2+r^2}(y\,dx-x\,dy)+\fracc{z}{r}dz\right]^2.
\end{array}
\end{equation}

The main characteristics of this coordinate system are:

\begin{itemize}

\item{For $M\rightarrow0$, it is Minkowski spacetime in Cartesian coordinates;}

\item{For $a\rightarrow0$, it is Schwarzschild solution in Cartesian coordinates;}

\item{The full metric ($M,a\neq0$) is obviously the Kerr-Schild form again.}

\end{itemize}

\item{Boyer-Lindquist coordinates:}

The most useful coordinate system for Kerr metric is Boyer-Lindquist coordinates, as follows

\begin{equation}
\label{ds2kerr_bl}
\begin{array}{l}
ds^2=\left(1-\fracc{2M\,r}{\rho^2}\right)d\hat t^2-\fracc{\rho^2}{\Delta}\,dr^2-\rho^2d\theta^2\\[2ex]
-\left(r^2+a^2+\fracc{2Ma\,r}{\rho^2}\,\sin^2\theta\right)\sin^2\theta\,d\hat\phi^2+\\[2ex]
+\fracc{4Ma\,r\,\sin^2\theta}{\rho^2}d\hat td\hat\phi,
\end{array}
\end{equation}
where $\Delta\equiv r^2+a^2-2Mr$.

The noticeable properties are that:

\begin{itemize}

\item{It minimizes the number of off-diagonal components of the metric;}

\item{The asymptotic behaviour of this coordinates permits to conclude that $M$ is indeed the mass and $J=Ma$ is the angular momentum;}

\item{For $a\rightarrow0$, it is Schwarzschild solution in standard coordinate system;}

\item{For $M\rightarrow0$, it is Minkowski line element in oblate spheroidal coordinates;}

\item{It is a maximal extension of the Kerr manifold.}

\end{itemize}

\item{Rational polynomial coordinates:}

If we make a coordinate transformation $\chi=\cos\theta$ from Boyer-Lindiquist coordinates, we have the following new version for Kerr spacetime

\begin{equation}
\label{ds2kerr_rp}
\begin{array}{l}
ds^2=\left(1-\fracc{2M\,r}{r^2+a^2\chi^2}\right)d\hat t^2-\fracc{r^2+a^2\chi^2}{\Delta}\,dr^2+\\[2ex]
-\fracc{(r^2+a^2\chi^2)}{1-\chi^2}d\chi^2+\fracc{4Ma\,r(1-\chi^2)}{r^2+a^2\chi^2}d\hat td\hat\phi+\\[2ex]
-(1-\chi^2)\left(r^2+a^2+\fracc{2Ma\,r(1-\chi^2)}{r^2+a^2\chi^2}\right)\,d\hat\phi^2.\\[2ex]
\end{array}
\end{equation}

These coordinates introduce the following qualities:

\begin{itemize}

\item{All metric components are rational polinomial of the coordinates;}

\item{The non-appearance of trigonometric functions do the computational calculations faster}

\end{itemize}

\item{Doran coordinates:}

Introduced by C. Doran in 2000, here we obtain another coordinate system for Kerr metric given by

\begin{equation}
\begin{array}{l}
\label{ds2kerr_doran}
ds^2=dt^2-\rho^2d\theta^2-(r^2+a^2)\sin^2\theta d\phi^2+\\[2ex]
-\fracc{\rho^2}{r^2+a^2}\left[dr+\fracc{\sqrt{2Mr(r^2+a^2)}}{\rho^2}(dt-a\sin^2\theta
d\phi)\right]^2.
\end{array}
\end{equation}

The useful features of Doran coordinates are:

\begin{itemize}

\item{For $a\rightarrow0$, it is Schwarzschild geometry in Painlevé-Gullstrand form;}

\item{For $M\rightarrow0$, we obtain Minkowski spacetime in oblate spheroidal coordinates;}

\item{In Doran coordinates, the contravariant metric component $g^{00}$ is equal to $1$;}

\item{According to ADM formalism, Doran coordinates slice the Kerr metric such that the ``lapse" function is everywhere unity.}

\end{itemize}

\item{Gaussian coordinates:}

Constructed from the relativistic Hamilton-Jacobi equation, The line element for the Kerr Metric in a Gaussian coordinate system is

\begin{equation}
\begin{array}{l}
\label{ds2kerr_gauss}
\nonumber
ds^2=dT^2-\left(E^2-1+\fracc{2Mr}{\rho^2}\right)dR^2+\\[2ex]
-\left(L^2+\tilde r^2\sin^2\theta+\fracc{2Mra^2\sin^4\theta}{\rho^2}\right)d\Theta^2+\\[2ex]
-4(r^2+\gamma)(\gamma-a^2\cos^2\theta)d\Phi^2+\\[2ex]
-\left(EL+\fracc{2Mar\sin^2\theta}{\rho^2}\right)dRd\Theta+\\[2ex]
-2\left[E(\gamma-a^2)+aL\right]dRd\Phi+\\[2ex]
-2\left[L(r^2+\gamma)-(E\tilde r^2-aL)a\sin^2\theta\right]d\Theta d\Phi,
\end{array}
\end{equation}
where $\tilde r=r^2+a^2$.

Besides the usual characteristics of the Gaussian coordinates, in this case we have that

\begin{itemize}

\item{For massive test particles, the geodesic equations parameterized by the proper time are immediately integrated;}

\item{Differently from all other cases, the metric is non-static but it is stationary;}

\item{This metric depends on parameters of the observers field comoving to the reference frame;}

\end{itemize}

\end{itemize}

\section{Schwarzschild solution from quasi-Maxwellian equations written in Gaussian coordinates}\label{appendixB}

As it was showed in \cite{lichnerowicz}, the JEK
(Jordan-Ehlers-Kundt) equations are equivalent to general
relativity. Besides, they become particularly simple when expressed
in a Gaussian coordinate system. Therefore, in principle, we could
use this simplicity to search of an internal solution for the Kerr
metric. As an example, we will apply this method to obtain the
stellar Schwarzschild solution.

There are many references treating the formal deduction of these
equations and their properties as for instance in \cite{dedqmax}.
These equations (JEK) can be obtained from Bianchi's identities,
together with Einstein equation, that is
\begin{equation}
\label{bianchi}
W^{\alpha\beta\mu\nu}{}_{;\nu}=-\fracc{1}{2}T^{\mu[\alpha;\beta]}+\fracc{1}{6}g^{\mu[\alpha}T^{,\beta]}
\end{equation}
From this we obtain the corresponding independent projections of the
divergence of Weyl tensor

\begin{equation}
\label{proj_div_weyl}
\begin{array}{l}
W^{\alpha\beta\mu\nu}{}_{;\nu}V_{\beta}V_{\mu}h_{\alpha}{}^{\sigma},\\
W^{\alpha\beta\mu\nu}{}_{;\nu}\eta^{\sigma\lambda}{}_{\alpha\beta}V_{\mu}V_{\lambda},\\
W^{\alpha\beta\mu\nu}{}_{;\nu}h_{\mu}{}^{(\sigma}\eta^{\tau)\lambda}{}_{\alpha\beta}V_{\lambda},\\
W^{\alpha\beta\mu\nu}{}_{;\nu}V_{\beta}h_{\mu(\tau}h_{\sigma)\alpha}.
\end{array}
\end{equation}

Besides, we also have the conservation law of the energy-momentum
tensor

\begin{equation}
\label{conserv_mom_eneg}
T^{\mu\nu}{}_{;\nu}=0,
\end{equation}
which can be projected parallel to an observer field $V^{\mu}$ or
perpendicular to it, as it follows

\begin{equation}
\label{proj_conserv_mom_eneg}
\begin{array}{l}
T^{\mu\nu}{}_{;\nu}V^{\mu}=0,\\
T^{\mu\nu}{}_{;\nu}h^{\mu}{}^{\alpha}=0,
\end{array}
\end{equation}
where $h^{\mu\nu}\doteq g^{\mu\nu}-V^{\mu}V^{\nu}$.

From the Riemann tensor, we can find the evolution equations for the kinematical quantities \textit{expansion} $\vartheta$, \textit{shear} $\sigma_{\mu\nu}$ and \textit{vorticity} $\omega_{\mu\nu}$ given by

\begin{equation}
\begin{array}{l}
\dot\vartheta+\fracc{\vartheta^2}{3}+2(\sigma^2+\omega^2)-a^{\alpha}{}_{;\alpha}=R_{\mu\nu}V^{\mu}V^{\nu},\\[1ex]
h_{\alpha}{}^{\mu}h_{\beta}{}^{\nu}\dot\sigma_{\mu\nu}+\fracc{1}{3}h_{\alpha\beta}(-2(\sigma^2+\omega^2)+a^{\lambda}{}_{;\lambda})+a_{\alpha}a_{\beta}+\\[1ex]
-\fracc{1}{2}h_{\alpha}{}^{\mu}h_{\beta}{}^{\nu}(a_{\mu;\nu}+a_{\nu;\mu})+\fracc{2}{3}\vartheta\sigma_{\alpha\beta}+\sigma_{\alpha\mu}\sigma^{\mu}{}_{\beta}+\\[1ex]
+\omega_{\alpha\mu}\omega^{\mu}{}_{\beta}=R_{\alpha\epsilon\beta\nu}V^{\epsilon}V^{\nu}-\fracc{1}{3}R_{\mu\nu}V^{\mu}V^{\nu}h_{\alpha\beta},\\[1ex]
h_{\alpha}{}^{\mu}h_{\beta}{}^{\nu}\dot\omega_{\mu\nu}-\fracc{1}{2}h_{\alpha}{}^{\mu}h_{\beta}{}^{\nu}(a_{\mu;\nu}-a_{\nu;\mu})+\fracc{2}{3}\vartheta\omega_{\alpha\beta}+\\[1ex]
-\sigma_{\beta\mu}\omega^{\mu}{}_{\alpha}+\sigma_{\alpha\mu}\omega^{\mu}{}_{\beta}=0,
\end{array}
\label{evol_quant_cine}
\end{equation}

together with the constraints

\begin{equation}
\label{eq_vinc_quant_cine}
\begin{array}{l}
\fracc{2}{3}\vartheta_{,\mu}h^{\mu}{}_{\lambda}-(\sigma^{\alpha}{}_{\gamma}+\omega^{\alpha}{}_{\gamma})_{;\alpha}h^{\gamma}{}_{\lambda}-a^{\nu}(\sigma_{\lambda\nu}+\omega_{\lambda\nu})=\\[1ex]
=R_{\mu\nu}V^{\mu}h^{\nu}{}_{\lambda},\\[1ex]
\omega^{\alpha}{}_{;\alpha}+2\omega^{\alpha}a_{\alpha}=0,\\[1ex]
-\fracc{1}{2}h_{(\tau}{}^{\epsilon}h_{\lambda)}{}^{\alpha}\eta_{\epsilon}{}^{\beta\gamma\nu}V_{\nu}(\sigma_{\alpha\beta}+\omega_{\alpha\beta})_{;\gamma}+a_{(\tau}\omega_{\lambda)}=H_{\tau\lambda}.
\end{array}
\end{equation}
If we assume that Einstein equation is only valid in a Cauchy surface, the set of equations\ (\ref{proj_div_weyl})-(\ref{eq_vinc_quant_cine}), so-called the \textit{quasi-Maxwellian equations}, propagates it to the whole spacetime.

Let us consider a diagonal metric, similar to Schwazschild one described in Gaussian coordinates, as follows

\begin{equation}
\label{metr_diag_max}
ds^2=dT^2-B(T,R)dR^2-r^2(T,R)d\Omega^2,
\end{equation}
and an observer field $V^{\mu}\doteq\delta^{\mu}_0$. The expansion $\vartheta$ for this vector is given by

\begin{equation}
\label{exp_diag_max} \vartheta=\fracc{1}{2}\left(\fracc{\dot B}{B}+\fracc{4\dot
r}{r}\right),
\end{equation}
where $\dot Y(T,R)\doteq\partial Y/\partial T$. After that, we calculate the shear tensor $\sigma^{\mu}{}_{\nu}$ and the eletric part of Weyl tensor ($E^{\mu}{}_{\nu}\doteq-W_{\alpha\mu\beta\nu}V^{\mu}V^{\nu}$) and write them in matricial form

\begin{equation}
\label{shear_diag_max}
[\sigma^i{}_j]=f(T,R)
\left(
\begin{array}{ccc}
1&0&0\\
0&-\fracc{1}{2}&0\\
0&0&-\fracc{1}{2}
\end{array}
\right),
\end{equation}

\begin{equation}
\label{elec_diag_max}
[E^i{}_j]=g(T,R)
\left(
\begin{array}{ccc}
1&0&0\\
0&-\fracc{1}{2}&0\\
0&0&-\fracc{1}{2}
\end{array}
\right),
\end{equation}
where

\begin{equation}
\label{f_diag_max} f(T,R)=\fracc{1}{3}\left(\fracc{\dot B}{B}-\fracc{2\dot r}{r}\right).
\end{equation}
and
\begin{equation}
\label{g_diag_max}
\begin{array}{l}
g(T,R)=\fracc{1}{12r^2B^2}(-2r^2B\ddot B+r^2\dot B^2-4rBr''+\\[1ex]
+2rB\dot r\dot B+2rr'B'+4rB^2\ddot r-4B^2-4B^2\dot r^2+\\[1ex]
+4Br'^2),
\end{array}
\end{equation}
where $Y'(T,R)\doteq\partial Y/\partial R$.

Observe that both $\sigma_{\mu\nu}$ and $E_{\mu\nu}$ are
proportional. All other quantities like the magnetic part of Weyl
tensor
($H_{\alpha\beta}\doteq-^{*}W_{\alpha\mu\beta\nu}V^{\mu}V^{\nu}$),
vorticity $\omega_{\alpha\beta}$ and acceleration ($a^{\mu}\doteq
V^{\mu}{}_{;\nu}V^{\nu}$) are identically zero, due to properties of
the observers congruence chosen.

Let us assume that $V^{\mu}=\delta^{\mu}_0$ in Gaussian coordinates
is co-moving to an arbitrary fluid, which can be expressed by

\begin{equation}
\label{ener_mom_diag_max}
T_{\mu\nu}=(\rho+p)V_{\mu}V_{\nu}-pg_{\mu\nu}+q_{(\mu}V_{\nu)}+\pi_{\mu\nu}
\end{equation}
where $\rho$ is the energy density, $p$ is the isotropic pressure,
$q_{\mu}$ is the heat flux and $\pi_{\mu\nu}$ is the anisotropic
pressure.

In the case of Schwarzschild stellar solution, it is assumed a
perfect fluid inside a spherical shell and an accelerated observer
($u_{\mu}=\sqrt{g_{00}}\,\delta_{\mu}^0$) comoving to this fluid.
With respect to the Gaussian observers such fluid presents a heat
flux $q^{\mu}=(0,q^1,0,0)$ and an anisotropic pressure

\begin{equation}
\label{ani_pres_s_g}
[\pi^i{}_j]=\pi(T,R)
\left(
\begin{array}{ccc}
1&0&0\\
0&-\fracc{1}{2}&0\\
0&0&-\fracc{1}{2}
\end{array}
\right).
\end{equation}

With these considerations, the set of equations\ (\ref{proj_div_weyl}) takes the form

\begin{equation}
\label{expl_qm_diag}
\begin{array}{lcl}
g'+3\fracc{r'}{r}g&=&\fracc{1}{3}\rho'+\fracc{\dot r}{r}q_1+\fracc{1}{2}\left(\pi'+3\fracc{r'}{r}\pi\right),\\[1ex]
\dot g+3\fracc{\dot
r}{r}g&=&\fracc{1}{4}f\pi-\fracc{1}{2}(\rho+p)f+\fracc{1}{2}\dot\pi+\fracc{1}{6}\vartheta\pi+\\[1ex]
&&-\fracc{1}{3}\left[(q^1)'+\left(\fracc{1}{2}\fracc{B'}{B}-2\fracc{r'}{r}\right)q^1\right],
\end{array}
\end{equation}
The conservation law\ (\ref{proj_conserv_mom_eneg}) can be written explicitly as

\begin{equation}
\label{expl_con_ener_max}
\begin{array}{l}
\dot\rho+(\rho+p)\vartheta-\fracc{3}{2}f\pi+(q^1)'+\left(\fracc{1}{2}\fracc{B'}{B}+2\fracc{r'}{r}\right)=0,\\
\pi'+3\fracc{r'}{r}\pi+q_{1,0}+\vartheta q_1-p'=0.
\end{array}
\end{equation}
Now, the evolution of the kinematical quantities is the following

\begin{equation}
\label{expl_evol_quant_cin_max}
\begin{array}{l}
\dot\vartheta+\fracc{\vartheta^2}{3}+\fracc{3}{2}f^2=-\fracc{1}{2}(\rho+3p),\\
\dot f+\fracc{f^2}{2}+\fracc{2}{3}\vartheta f=-g-\fracc{1}{2}\pi^1{}_1.
\end{array}
\end{equation}
Finally, the only constraint equation is

\begin{equation}
\label{expl_vin_max} f'+3\fracc{r'}{r}f-\fracc{2}{3}\vartheta'=0.
\end{equation}

Hereupon the set of equations\
(\ref{expl_qm_diag})-(\ref{expl_vin_max}) corresponds to a problem
of initial conditions, which shall give origin to Schwarzschild
solution. As it is not our aim, we will not solve these equations
step by step. However, we will indicate how to proceed.

First, we analyze Einstein equations for $T_{\mu\nu}=0$ and we
obviously obtain $B(T,R)$ and $r(T,R)$ like those given in Eq.\
(\ref{ds2_schwarz_mario_gauss}), identifying $T=\tau$. On the other
hand, from the quasi-Maxwellian equations we get

\begin{equation}
\label{eq_func_schwarz}
\begin{array}{l}
B=\fracc{r'^2}{1+h(R)},\\[2ex]
\dot r=\sqrt{y(R)+\fracc{k}{r}},\\[2ex]
r'=b(R)\sqrt{h(R)+\fracc{k}{r}}
\end{array}
\end{equation}
where $h(R)$, $y(R)$ and $b(R)$ are arbitrary functions and $k$ is a
constant. If we assume as initial condition surface $r(T,R)\equiv
const\rightarrow\infty$, then we will obtain the Schwarzschild
external solution.

The Schwarzschild internal solution can be obtained if we consider
the energy-momentum tensor associated to the Gaussian observer
$\delta^{\mu}_0$, written in terms of the quantities associated to
the observer $u_{\mu}$ (energy density $\rho$ and pressure $p$)
presented in Sec.\ [\ref{schwar_int_case}], as follows

\begin{equation}
\label{ener_mom_gauss}
\begin{array}{l}
\rho_G=(\rho+p)\alpha^2e^{-\nu}-p\\[2ex]
p_G=-\fracc{1}{3}[(\rho+p)(1-\alpha^2e^{-\nu})-3p]\\[2ex]
q^{\mu}=(\rho+p)\alpha e^{-\nu}(0,1,0,0)\\[2ex]
\pi=\fracc{2}{3}(1-\alpha^2e^{-\nu})
\end{array}
\end{equation}
where $\nu=\nu(T,R)$ and $\alpha$ is an external parameter.
Substituting these equations in the quasi-Maxwellian equations, we
will find exactly the Schwarzschild internal solution with some
arbitrary functions. However, we must match this solution with that
coming from the initial condition, that is Einstein equation on the
hyper-surface. Besides, choosing as Cauchy surface $r(T,R)=r_0$, we
 fix the arbitrary functions, obtaining with such a procedure the Schwarzschild
stellar solution.

From this example, we conclude that the search of an internal
solution for the Kerr metric most naturally should not be reduced to
a perfect fluid in the Gaussian coordinate system. We should expect
that the associated observer detects a heat flux, as in the case of
static spherically symmetric, as above. It is clear that the
quasi-Maxwellian equations for the Kerr metric in the Gaussian
system is rather more involved than in the Schwarzschild case. So,
as future work, we intend to modify such Gaussian coordinate system
found, making some spatial coordinate transformations and then, to
find an internal solution for Kerr metric using complex fluids but
Gaussian observers.

\end{document}